\documentclass[11pt]{article}

\usepackage{amsmath,amssymb,theorem}
\usepackage{amscd}

\numberwithin{equation}{section}

\theorembodyfont{\rmfamily}

\newcommand{\sfrac}[2]{{\textstyle\frac{#1}{#2}}}
\newcommand{\ns}{\normalsize}

\newcommand{\hf}{\frac{1}{2}}
\newcommand{\shf}{\sfrac{1}{2}}

\newcommand{\ZZ}{{\mathbb{Z}}}
\newcommand{\PP}{{\mathbb{CP}}}

\newcommand{\cE}{{\mathcal{E}}}

\newcommand{\cS}{{\mathcal{S}}}

\def\k{\kappa}
\def\l{\lambda}

\def\s{\sigma}
\def\t{\tau}

\def\D{\Delta}

\begin{document}


\begin{titlepage}

\title{
   \vspace*{-3em}   
   \hfill{\ns UPR-872T} \\[3em]
   {\LARGE Standard Model Vacua in Heterotic M--Theory}
       \\[1em] } 
\author
   {\ns Ron Donagi$^1$, Burt A.~Ovrut$^2$\setcounter{footnote}{0}\thanks
   {Invited talk at STRINGS'99, Potsdam, Germany, July 19-24, 1999.}~~, 
   Tony Pantev$^1$ and Daniel Waldram$^3$ \\ [0.5em]
   {\it\ns $^1$Department of Mathematics, University of Pennsylvania} \\[-0.3em]
   {\it\ns Philadelphia, PA 19104--6395, USA}\\ 
   {\it\ns $^2$Department of Physics, University of Pennsylvania} \\[-0.3em]
   {\it\ns Philadelphia, PA 19104--6396, USA}\\ 
   {\it\ns $^3$Theory Division, CERN CH-1211,} \\[-0.3em]
   {\it\ns Geneva 23, Switzerland}}
\date{}

\maketitle

\vfill

\begin{abstract}
We present a class of $N=1$ supersymmetric ``standard'' models 
of particle physics, derived
directly from heterotic M--theory, that contain three families of
chiral quarks and leptons coupled to the gauge group $SU(3)_{C} \times
SU(2)_{L} \times U(1)_{Y}$. These models are a fundamental form of
``brane world'' theories, with an observable and hidden sector each
confined, after compactification on a Calabi--Yau threefold, to a BPS
three-brane separated by a higher dimensional bulk space with size of
the order of the intermediate scale. The requirement of three
families, coupled to the fundamental conditions of anomaly freedom and
supersymmetry,
constrains these models to contain additional five-branes located in the bulk
space and wrapped
around holomorphic curves in the Calabi--Yau threefold. 
\end{abstract}

\vfill

\thispagestyle{empty}

\end{titlepage}


\section{Introduction}

In fundamental work, it was shown by Ho\v rava and Witten \cite{HW1,HW2} that if
M--theory is compactified on the orbifold $S^1/Z_2$,
a chiral $N=1$, $E_{8}$ gauge supermultiplet must exist in the twisted 
sector of each of the two ten-dimensional orbifold fixed planes.
It is important to note that, in this theory, the chiral gauge matter is
confined solely to the orbifold planes, while pure supergravity
inhabits the bulk space between these planes. Thus, Ho\v rava-Witten
theory is a concrete and fundamental representation of the idea of a
``brane world''. 

Witten then showed \cite{W} that, if further compactified to four dimensions on
a Calabi--Yau threefold, the $N=1$ supersymmetric low--energy theory
exhibits realistic gauge unification and gravitational coupling
strength provided the Calabi--Yau radius, $R$, is of the order of
$10^{16}$GeV and that the orbifold radius, $\rho$, is larger than
$R$. Thus, Ho\v rava--Witten theory has a ``large'' internal
bulk dimension, although it is of order the intermediate scale and not
the TeV size bulk dimensions, or larger, discussed recently \cite{dim}.

When compactifying the Ho\v rava--Witten theory, it is possible that
all or, more likely, a subset of the $E_{8}$ gauge fields do not
vanish classically in the Calabi--Yau threefold directions. Since
these gauge fields ``live'' on the Calabi--Yau manifold,
$3+1$-dimensional Lorentz invariance is left unbroken. Furthermore, by
demanding that the associated field strengths satisfy the constraints
$F_{ab}=F_{\bar a\bar b}=g^{a \bar b}F_{a \bar b}=0$, $N=1$
supersymmetry is preserved. However, these gauge field vacua do
spontaneously break the $E_{8}$ gauge group as follows. Suppose that
the non-vanishing gauge fields are associated with the generators of a
subgroup $G$, where $G\times H \subseteq E_{8}$. Then the $E_{8}$
gauge group is spontaneously broken to $H$, which is the commutant
subgroup of $G$ in $E_{8}$. This mechanism of gauge group breaking
allows one, in principle, to reduce the $E_{8}$ gauge group to smaller
and phenomenologically more interesting gauge groups such as
unification groups $E_{6}$, $SO(10)$ and $SU(5)$ as well as the
standard model gauge group $SU(3)_{C} \times SU(2)_{L} \times
U(1)_{Y}$. The spontaneous breaking of $E_{8}$ to $E_{6}$ by taking
$G=SU(3)$ and identifying it with the spin connection of the Calabi--Yau
threefold, the so-called ``standard embedding'', was discussed in
\cite{HW1,HW2}. A general discussion of non-standard embeddings in this
context and their low energy implications was presented in \cite{nse,lpt}. We will
refer to Ho\v rava--Witten theory compactified to lower dimensions with
arbitrary gauge vacua as heterotic M--theory.

It is, therefore, of fundamental interest to know, given a Calabi--Yau
threefold $X$, what non-Abelian gauge field vacuum configurations
associated with a subgroup $G\subseteq E_{8}$ can be defined on
it. One approach to this problem is to simply attempt to solve the
six-dimensional Yang--Mills equations with the appropriate boundary
conditions subject to the above constraints on the field
strengths. However, given
the complexity of Calabi--Yau threefolds, this approach becomes very
difficult at best and is probably untenable. One, therefore, must look
for an alternative construction of these Yang-Mills connections. Such
an alternative was presented by Donaldson \cite{Don} and Uhlenbeck and Yau
\cite{UhYau}, who
recast the problem in terms of holomorphic vector bundles. These
authors prove that for every semi-stable holomorphic vector bundle
with structure group $G$ over $X$, there exists a solution to the
six-dimensional Yang--Mills equations satisfying the above constraints
on the field strengths, and conversely. Thus, the problem of
determining the allowed gauge vacua on a Calabi--Yau threefold is
replaced by the problem of constructing semi--stable holomorphic
vector bundles over the same threefold.  

It was shown in recent publications
\cite{don1,don2,don3,curio,ba},
relying heavily on work on holomorphic vector bundles by several authors
\cite{FMW,D,BJPS}, that a wide class of semi-stable holomorphic vector bundles
with structure groups $SU(n)\subset E_{8}$ can be explicitly
constructed over elliptically fibered Calabi--Yau threefolds. The
restriction to $SU(n)$ subgroups was for simplicity, other structure
subgroups being possible as well. Thus, using holomorphic vector
bundles and the Donaldson, Uhlenbeck, Yau theorem, it has been possible
to classify and give the properties of a large class of $SU(n)$ gauge
vacua even though the associated solutions of the Yang--Mills
equations are unknown. 

As presented in~\cite{don1,don2}, three--family vacua with
phenomenologically interesting unification groups such as $E_{6}$,
$SO(10)$ and $SU(5)$ could be obtained, corresponding to vector bundle
structure groups $SU(3)$, $SU(4)$ and $SU(5)$ respectively. However,
it was not possible to break $E_{8}$ directly to the standard gauge
group $SU(3)_{C} \times SU(2)_{L} \times U(1)_{Y}$ in this manner. A
natural solution to this problem is to use non-trivial Wilson lines to
break the GUT group down to the standard gauge group \cite{bos,ow}. This requires
that the fundamental group of the Calabi--Yau threefold be
non-trivial. Unfortunately, one can show that all elliptically fibered
Calabi--Yau threefolds are simply connected, with the exception of
such threefolds over an Enriques base which, however \cite{don2}, 
is not consistent with the requirement of three
families of quarks and leptons. 

With this in mind, recall that an elliptic fibration is simply a torus
fibration that admits a zero section. We were able to show that it is
the requirement of a zero section that severely restricts the
fundamental group of the threefold to be, modulo the one exception
mentioned above, trivial. Hence, if one lifts the zero section
requirment, and considers holomorphic vector bundles over
torus-fibered Calabi--Yau threefolds without section, then one expects
to find non-trivial first homotopy groups and Wilson lines in vacua
that are consistent with the three-family requirement. In \cite{me} we gave
the relevant mathematical properties of
a specific class of torus-fibered Calabi--Yau threefolds without
section and constructed holomorphic vector bundles  over such
threefolds. We then used these results to explicitly construct a number
of three-family vacua with unification group $SU(5)$ which is
spontaneously broken to the standard gauge group $SU(3)_{C} \times
SU(2)_{L} \times U(1)_{Y}$ by Wilson lines. 

The results of \cite{me} represent $N=1$ ``standard''
models of particle physics derived directly from M--theory. Each of
these vacua has three families of chiral quarks and leptons coupled
to the standard $SU(3)_{C} \times SU(2)_{L} \times U(1)_{Y}$ gauge
group. As discussed above, this ``observable sector'' lives on a $3+1$
dimensional ``brane world''. It was shown in~\cite{losw1,losw2} that this
$3+1$ dimensional space is the worldvolume of a BPS three--brane. It
is separated from a ``hidden sector'' three--brane by a bulk space
with an intermediate scale ``large'' extra dimension. The requirement
of three families, coupled to the fundamental condition of anomaly
freedom and supersymmetry, constrains the theory to admit 
an effective class describing
the wrapping of additional five-branes around holomorphic curves in
the Calabi--Yau threefold. These five-branes ``live'' in the bulk
space and represent new, non-perturbative aspects of particle physics
vacua. 

In this talk, we present the rules for building phenomenological particle
physics ``standard'' models in heterotic M-theory on torus-fibered
Calabi--Yau threefolds without section realized as quotient manifolds
$Z=X/\tau_{X}$. These quotient threefolds have a non-trivial first homotopy group
$\pi_{1}(Z)=\ZZ_{2}$.
Specifically, we construct three-family
particle physics vacua with GUT group $SU(5)$. Since $\pi_{1}(Z)=\ZZ_{2}$, these
vacua have Wilson lines that break $SU(5)$ to the standard $SU(3)_{C}
\times SU(2)_{L} \times U(1)_{Y}$ gauge group. 
We then present several explicit examples of these ``standard'' model vacua for the
base surface $B=F_{2}$ of the torus fibration. We refer the reader to \cite{me} for
the mathematical details and a wider set of examples, including the base $B=dP_{3}$.


\section{Rules for Realistic Particle Physics Vacua}

In this section, we give the rules required to construct realistic 
particle physics vacua, restricting our results to vector bundles with
structure group $SU(n)$ for $n$ odd. The rules presented here lead 
to $N=1$ supersymmetric
theories with three families of quarks and leptons with the standard
model gauge group $SU(3)_{C}\times SU(2)_{L}\times U(1)_{Y}$. 

The first set of rules deals with the selection of the
elliptically fibered Calabi--Yau threefold $X$ with two sections, the
choice of the involution and constraints on the vector bundles, such that
the bundles descend to vector bundles on $Z=X/\t_X$. If one was
using this construction to construct vector bundles for each of the
two $E_8$ groups in Ho\v rava-Witten theory, then this first set of
constraints is applicable to each bundle individually. The rules are
\begin{itemize}
\item 
Two Section Condition: Choose an elliptically fibered Calabi--Yau
threefold $X$ which admits two sections $\sigma$ and $\xi$. This is
done by selecting the base manifold $B$ of $X$ to be a 1) del Pezzo,
2) Hirzebruch, 3) blown-up Hirzebruch or 4) an Enriques surface. The
threefold $X$ with two sections is then specified by its Weierstrass
model with an explicit choice of 
\begin{equation}
   g_{2} = 4(a^{2}-b) , \qquad 
   g_{3} = 4ab .
\label{eq:43}
\end{equation}
The discriminant is then given by
\begin{equation}
   \Delta = \Delta_{1}\Delta_{2}^{2} ,
\label{eq:44}
\end{equation}
where
\begin{equation}
   \Delta_{1} = a^{2}-4b , \qquad  
   \Delta_{2}=4(2a^{2}+b) .
\label{eq:45}
\end{equation}
\item 
Choice of Involution: Using the properties of the base, explicitly
specify an involution $\tau_{B}$ on $B$. 
Now choose sections $a$ and $b$ to be invariant under $\tau_{B}$. This allows
one to construct an involution $\tau_{X}$ on $X$.
Find the set of fixed points
$\mathcal{F}_{\t_B}$ under $\tau_{B}$ and show that 
\begin{equation}
   \mathcal{F}_{\t_B} \cap \{\Delta=0\} = \emptyset .
\label{eq:46}
\end{equation}
\item 
Bundle Constraint: Consider semi-stable holomorphic vector bundles $V$ over
$X$. To construct any such vector bundle one must specify a divisor
class $\eta$ in the base $B$ as well as coefficients $\lambda$ and
$\kappa_i$. These coefficients satisfy 
\begin{equation}
   \l - \sfrac{1}{2} \in \ZZ , \qquad
   \k_i - \shf m \in \ZZ , 
\label{eq:47}
\end{equation}
with $m$ an integer. Furthermore, we must have that
\begin{equation}
\eta \text{ is effective}
\label{eq:47A}
\end{equation}
as a class on $B$.

\item 
Bundle Involution Condition: In order for $V$ to descend to a vector
bundle $V_{Z}$ over $Z$, the class $\eta$ in $B$ and the coefficients
$\k_i$ must satisfy the constraints
\begin{equation}
\begin{aligned}
   \tau_{B}(\eta) &= \eta , \\
   \sum_i \k_i &= \eta\cdot c_1
\end{aligned}
\label{eq:48}
\end{equation}
\end{itemize}

The second set of rules is directly particle physics related. The 
first of these is the requirement that the theory have three families of 
quarks and leptons. The number of generations associated with the vector 
bundle $V_{Z}$ over $Z$ is given by
\begin{equation}
   N_{\text{gen}} = \sfrac{1}{2}c_{3}(V_{Z}) . 
\label{eq:49}
\end{equation}
Requiring $N_{\text{gen}}=3$ leads to the following rule for the associated vector
bundle $V$ over $X$. 
\begin{itemize}
\item 
Three-Family Condition: To have three families we must require
\begin{equation}
   6 = \lambda \eta( \eta-nc_{1}) .
\label{eq:50}
\end{equation}
\end{itemize}

The second such rule is associated with the anomaly cancellation
requirement that 
\begin{equation}
   [W_{Z}] = c_{2}(TZ) - c_{2}(V_{Z1}) - c_{2}(V_{Z2}) ,
\label{eq:51}
\end{equation}
where $[W_{Z}]$ is the class associated with non-perturbative
five-branes in the bulk space of the Ho\v rava-Witten theory. Vector
bundles $V_{Z1}$ and $V_{Z2}$ are located on the ``observable'' and
``hidden'' orbifold planes respectively. In this talk, for
simplicity, we will always take $V_{Z2}$ to be the trivial
bundle. Hence, gauge group $E_{8}$ remains unbroken on the ``hidden''
sector, $c_{2}(V_{Z2})$ vanishes and condition \eqref{eq:51}
simplifies accordingly. Using the definition 
\begin{equation}
   [W_{Z}] = \frac{1}{2}q_{*}[W] ,
\label{eq:51A}
\end{equation}
condition \eqref{eq:51} can be pulled-back onto $X$ to give
\begin{equation}
   [W] = c_{2}(TX) - c_{2}(V) .
\label{eq:52}
\end{equation}
It follows that
\begin{equation}
   [W] = \sigma_{*}W_{B} + c(F-N) + dN
\label{eq:53}
\end{equation}
where
\begin{equation}
   W_{B} = 12c_1-\eta
\label{eq:54}
\end{equation}
and
\begin{align}
   c &= c_2 + \left(\frac{1}{24}(n^{3}-n)+11\right)c_1^2 
         - \hf\left(\l^2-\frac{1}{4}\right)
              n\eta\left(\eta-nc_1\right)
         - \sum_i\k_i^2  , 
\label{eq:55} \\
   d &= c_2 + \left(\frac{1}{24}(n^{3}-n)-1\right)c_1^2 
         - \hf\left(\l^2-\frac{1}{4}\right)
              n\eta\left(\eta-nc_1\right)
         - \sum_i\k_i^2 + \sum_i\k_i .
\label{eq:56}
\end{align}
The class $[W_{Z}]$ must represent an actual physical holomorphic
curve in the Calabi--Yau threefold $Z$ since physical five-branes are
required to wrap around it. Hence, $[W_{Z}]$ must be an effective
class and, hence, its pull-back $[W]$ is an effective class in the covering
threefold $X$. This leads to the following rule.
\begin{itemize}
\item 
Effectiveness Condition: For $[W]$ to be an effective class, we 
require
\begin{equation}
   W_{B} \text{ is effective in $B$}, \quad c \geq 0, \quad d \geq 0 .
\label{eq:57}
\end{equation}
\end{itemize}

Finally, consider subgroups of $E_{8}$ of the form
\begin{equation}
   G \times H \subset  E_{8} .
\label{eq:58}
\end{equation}
If $G$ is chosen to be the structure group of the vector bundle, then,
naively, one would expect the commutant subgroup $H$ to be the
subgroup preserved by the bundle. However, Rajesh, Berglund and Mayr 
\cite{R} have shown that
this will be the case if and only if the vector bundle satisfies a
further constraint. If this constraint is not satisfied, then the
actual preserved subgroup of $E_{8}$ will be larger than $H$. Although
not strictly necessary, we find it convenient in model building to
demand that this constraint hold. 
\begin{itemize}
\item 
Stability Constraint: Let $G \times H \subset E_{8}$ and $G$ be the
structure group of the vector bundle. Then $H$ will be the largest
subgroup preserved by the bundle if and only if
\begin{equation}
   \eta > nc_{1}. 
\label{eq:59}
\end{equation}
\end{itemize}

If one follows the above rules, then the vacua will correspond to a
grand unified theory with unification group $H$ and three families of
quarks and leptons. In this talk, we will only
consider the maximal subgroup $SU(5)\times SU(5)\subset E_{8}$. We
then choose 
\begin{equation}
   G = SU(5) . 
\label{eq:60}
\end{equation}
Therefore, the unification group will be 
\begin{equation}
   H = SU(5) . 
\label{eq:61}
\end{equation}
However, these vacua correspond to vector bundles over the quotient
torus-fibered Calabi--Yau threefold $Z$ which has non-trivial homotopy
group
\begin{equation}
   \pi_{1}(Z)=\ZZ_{2} . 
\label{eq:62}
\end{equation}
It follows that the GUT group will be spontaneously broken to the 
standard model gauge group 
\begin{equation}
   SU(5) \rightarrow SU(3)_{C} \times SU(2)_{L} \times U(1)_{Y} ,
\label{eq:63}
\end{equation}
if we adopt the following rule.
\begin{itemize}
\item 
Standard Gauge Group Condition: Assume that the bundle contains a
non-vanishing Wilson line with generator 
\begin{equation}
   \mathcal{G} = \left(\begin{array}{cc}
        \mathbf{1}_3 & \\ 
        & -\mathbf{1}_2 
        \end{array}\right) .
\label{eq:64}
\end{equation}
\end{itemize}

Armed with the above rules, we now turn to the explicit construction
of phenomenologically relevant non-perturbative vacua. 


\section{Three Family Models}

We begin by choosing the base of the Calabi--Yau threefold to be the
Hirzebruch surface
\begin{equation}
   B = F_2.
\end{equation}
As discussed in the Appendix of~\cite{don2}, the Hirzebruch surfaces
are $\PP^1$ fibrations over $\PP^1$. There are two independent classes
on $F_2$, the class of the base $\cS$ and of the fiber
$\cE$. Their intersection numbers are 
\begin{equation}
   \cS\cdot\cS = -2, \qquad
   \cS\cdot\cE = 1, \qquad
   \cE\cdot\cE = 0.
\label{eq:intc2}
\end{equation}
The first and second Chern classes of $F_2$ are given by 
\begin{equation}
   c_1(F_2) = 2\cS + 4\cE, 
\label{c1F2}
\end{equation}
and
\begin{equation}
   c_2(F_2) = 4.
\label{c2F2}
\end{equation}

We now need to specify the involution $\t_B$ on the base and how it
acts on the classes on $B$. We recall that there is a single type of
involution on $\PP^1$. If $(u,v)$ are homogenous coordinates on
$\PP^1$, the involution can be written as $(u,v)\to(-u,v)$. This
clearly has two fixed points, namely the origin $(0,1)$ and the point
at infinity $(1,0)$ in the $u$-plane. To construct the involution
$\t_B$, we combine an involution on the base $\PP^1$ with one on the
fiber $\PP^1$. Thus ${\cal{F}}_{\tau_{B}}$ contains four fixed points.

To ensure that we can construct a freely acting involution $\t_X$ from
$\t_B$, we need to show that the discriminant curve can be chosen so as not to
intersect these fixed points. We recall that the two components of the
discriminant curve are given by 
\begin{equation}
   \D_1 = a^2 - 4b, \qquad
   \D_2 = 4\left(2a^2 + b\right),
\end{equation}
and that parameters $a$ and $b$ are sections of $K_B^{-2}$ and $K_B^{-4}$
respectively, where $K_B$ is the canonical bundle of the base.
In order to lift $\t_B$ to an involution of $X$, we required that
\begin{equation}
   \t_B(a) = a, \qquad
   \t_B(b) = b.
\label{eq:hi} 
\end{equation}
This restricts the allowed sections $a$ and $b$ and, consequently, the
form of $\Delta_{1}$ and $\Delta_{2}$. One can show that, 
for a generic choice of $a$ and $b$ satisfying
conditions \eqref{eq:hi}, there is enough freedom so that the
discriminant curves do not intersect any of the fixed points. 

We now want to consider curves $\eta$ in $F_{2}$ that are invariant under the
involution $\tau_{B}$. This can be done by first determining how this
involution acts on the effective classes.
We find that the involution preserves both $\cS$ and $\cE$ separately, so that 
\begin{equation}
   \t_B(\cS) = \cS, \qquad
   \t_B(\cE) = \cE.
\label{eq:hello}
\end{equation}
Since any class $\eta$ is a linear combination of $\cS$ and $\cE$, we see that
an arbitrary $\eta$ satisfies $\tau_{B}(\eta)=\eta$.

We can now search for $\eta$, $\l$ and $\k_{i}$ satisfying the three
family, effectiveness and stability conditions given above. We find that
there are two classes of solutions
\begin{equation}
\begin{aligned}
   \text{solution 1:} & \quad 
       \eta = 14\mathcal{S} + 22\mathcal{E}, \quad 
       \l = \sfrac{3}{2}, \\
       {}& \sum_i\k_i = \eta\cdot c_1 = 44, \quad
       \sum_i \k_i^2 \leq 60 , \\
   \text{solution 2:} & \quad
       \eta = 24\mathcal{S} + 30\mathcal{E}, \quad 
       \l = -\sfrac{1}{2}, \\
       {}& \sum_i\k_i = \eta\cdot c_1 = 60, \quad
       \sum_i \k_i^2 \leq 76 .  
\end{aligned}
\label{solF2}
\end{equation}

First note that the coefficients $\l$ satisfy the bundle constraint
\eqref{eq:47}. Furthermore, one can find many examples of $\k_i$ with
$i=1,\dots,4\eta\cdot c_1$, satisfying the bundle
constraint~\eqref{eq:47}, the given conditions on $\sum_i\k_i^2$ and
the invariance condition $\sum_i\k_i=\eta\cdot c_1$. 

Using $n=5$, \eqref{c1F2}, \eqref{solF2} and the intersection
relations \eqref{eq:intc2}, one can easily verify that both
solutions satisfy the three-family condition \eqref{eq:50}. 

Next, from \eqref{eq:53}, \eqref{eq:54}, \eqref{eq:55}
and \eqref{eq:56}, as well as $n=5$, \eqref{c1F2}, \eqref{c2F2},
\eqref{solF2} and the intersection relations \eqref{eq:intc2}, we can
calculate the five-brane curves $W$ associated with each of the 
solutions. We find that 
\begin{equation}
\begin{aligned}
   \text{solution 1:} \quad & 
      [W] = \s_{*}\left(10\cS+26\cE\right)
             + \left(112-k\right)\left(F-N\right) 
             + \left(60-k\right) N, \\   
   \text{solution 2:} \quad & 
      [W] = \s_{*}\left(18\cE \right)
             + \left(132-k\right)\left(F-N\right) 
             + \left(76-k\right) N, 
\end{aligned}
\label{eq:branes}
\end{equation}
where
\begin{equation}
   k = \sum_i \k_i^2
\end{equation}
It follows that the base components for $[W]$ are given by
\begin{equation}
\begin{aligned}
   \text{solution 1:} \quad & W_{B} = 10\cS + 26\cE, \\
   \text{solution 2:} \quad & W_{B} = 18\cE,
\end{aligned}
\end{equation}
which are both effective. Furthermore, we note that for each
five-brane curve the $c$ and $d$ coefficients of classes $F-N$ and $N$
respectively are non-negative integers (given the constraints on
$k$). Hence, effectiveness condition \eqref{eq:57} is satisfied. 

Finally, note that for $n=5$ the stability condition becomes
$\eta > 5c_{1}$. In both of the above solutions  
\begin{equation}
   \eta > 5c_1 = 10\cS +20\cE
\end{equation}
so that the stability condition is satisfied. Note that this condition is
consistent with the somewhat stronger condition used in \cite{me} since $\eta$
and $c_{1}$ have integer coefficients.

We conclude that, over a Hirzebruch base $B=F_{2}$, one can construct
torus-fibered Calabi--Yau threefolds, $Z$, without section with
non-trivial first homotopy group $\pi_{1}(Z)=\ZZ_{2}$. Assuming a
trivial gauge vacuum on the hidden brane, we have shown that we expect
these threefolds to admit precisely two classes of semi-stable holomorphic
vector bundles $V_{Z}$, \eqref{solF2}, associated with an $N=1$
supersymmetric theory  with three families of chiral quarks and
leptons and GUT group $H=SU(5)$ on the observable brane world. Since
$\pi_{1}(Z)=\ZZ_{2}$, Wilson lines break this GUT group as 
\begin{equation}
   SU(5) \rightarrow SU(3)_{C} \times SU(2)_{L} \times U(1)_{Y} ,
\end{equation}
to the standard model gauge group. Anomaly cancellation and
supersymmetry require the existence of non-perturbative five-branes in
the extra dimension of the bulk space. These five-branes are wrapped
on holomorphic curves in $Z$ whose homology classes,
\eqref{eq:branes}, are exactly calculable.

\subsection*{Acknowledgments}

R. Donagi is supported in part by an NSF grant DMS-9802456 as well as a 
University of Pennsylvania Research Foundation Grant. 
B.A. Ovrut is supported in part by a Senior Alexander von Humboldt Award, 
by the DOE under contract No. DE-AC02-76-ER-03071 and by a University 
of Pennsylvania Research Foundation Grant.
T. Pantev is supported in part by an NSF grant DMS-9800790 and by an Alfred P.
Sloan Research Fellowship.




\begin{thebibliography}{99}

\bibitem{HW1} P. Ho\v rava and E. Witten, {\em Nucl. Phys.} {\bf B460}
    (1996) 506.
\bibitem{HW2} P. Ho\v rava and E. Witten, {\em Nucl. Phys.} {\bf B475}
    (1996) 94.
\bibitem{W} E. Witten, {\em Nucl. Phys.} {\bf B471} (1996) 135.
\bibitem{dim} N. Arkani-Hamed, S. Dimopoulos and G. Dvali, 
    {\em Phys. Lett.} {\bf B429} (1998) 263; 
    I. Antoniadis, N. Arkani-Hamed, S. Dimopoulos and G. Dvali, 
    {\em Phys. Lett.} {\bf B436} (1998) 257; 
    L. Randall and R. Sundrum, {\em An Alternative to
    Compactification},  hep-th/9906064;
    K. R. Dienes, E. Dudas and T. Gherghetta, 
    {\em Phys. Lett.}{\bf B436} (1998) 55. 
\bibitem{nse} A. Lukas, B.~A. Ovrut and D. Waldram, 
    {\em Phys.Rev.} {\bf D59} (1999) 106005.
\bibitem{lpt} Z. Lalak, S. Pokorski and S. Thomas, 
    {\em Nucl. Phys.} {\bf B549}, (1999) 63.
\bibitem{Don} S. Donaldson, {\em Proc. London Math. Soc.} {\bf 3} 1 (1985).
\bibitem{UhYau} K. Uhlenbeck and S.-T. Yau, {Comm. Pure App. Math.}{\bf 39}
    257 (1986), {\bf 42} 703 (1986).
\bibitem{don1} R. Donagi, A. Lukas, B.~A. Ovrut and D. Waldram, 
    {\em JHEP} {\bf 05} (1999) 018.
\bibitem{don2} R. Donagi, A. Lukas, B.A. Ovrut, D. Waldram, 
    {\em JHEP} {\bf 06} (1999) 034. 
\bibitem{don3} R. Donagi, B.A. Ovrut and D. Waldram, {\em  Moduli Spaces of 
    Fivebranes on Elliptic Calabi-Yau Threefolds},  hep-th/9904054.
\bibitem{curio} G. Curio, {\em Phys. Lett.} {\bf B435} 39 (1998).
\bibitem{ba} B. Andreas, 
    {\em JHEP} {\bf 01} (1999) 011.
\bibitem{FMW} R. Friedman, J. Morgan and E. Witten,
    \textit{Commun. Math. Phys.} \textbf{187}, 679 (1997).
\bibitem{D} R. Donagi, {\em Asian. J. Math.} {\bf 1}, 214 (1997).
\bibitem{BJPS} M. Bershadsky, A. Johansen, T. Pantev and V. Sadov, 
    {\em Nuc. Phys.} {\bf B505} 165 (1997).
\bibitem{bos} J.D. Breit, B.A. Ovrut, G.C. Segre, 
    {\em Phys. Lett.} {\bf 158B} (1985) 33.
\bibitem{ow} E. Witten, 
    {\em Nucl. Phys.} {\bf B258} (1985) 75.
\bibitem{me} R. Donagi, B.A. Ovrut, T. Pantev and D. Waldram, hep-th/9912208.
\bibitem{losw1} A. Lukas, B.~A. Ovrut, K.~S. Stelle and D. Waldram,
    {\em  Phys. Rev.} {\bf D59} (1999) 086001.
\bibitem{losw2} A. Lukas, B.~A. Ovrut, K.~S. Stelle and D. Waldram,
    {\em Nucl. Phys} {\bf B552} (1999) 246.
\bibitem{R} G. Rajesh, 
   {\em JHEP} {\bf 12} (1998) 018;
   P. Berglund and P. Mayr,
   {\em Adv. Theor. Math. Phys.} {\bf 2} (1999) 1307-1372.
\end{thebibliography}
\end{document}